\documentclass[preprint,12pt]{elsarticle}

\usepackage{graphicx}
\usepackage{color}
\usepackage{url}
\usepackage{amsmath}
\usepackage{amssymb}
\usepackage{dcolumn}

\newcommand{\s}{\sum}
\newcommand{\f}{\frac}
\newcommand{\rro}{\right)}
\newcommand{\lro}{\left( }
\newcommand{\lsq}{\left[}
\newcommand{\rsq}{\right]}

\newcommand{\beq}{\begin{equation}}
\newcommand{\eeq}{\end{equation}}
\newcommand{\Teleia}{\quad .}
\newcommand{\Comma}{\quad ,}

\newcommand{\esp}{ESPResSo}
\newcommand{\bpar}{\beta_{k}^{\parallel}}
\newcommand{\bperp}{\beta_{k}^{\perp}}

\biboptions{comma,square}

\journal{Journal of Colloid and Interface Science}

\begin{document}

\begin{frontmatter}

\title{On the friction coefficient of straight-chain aggregates}

\author[jrc]{Lorenzo Isella\fnref{present}}
\ead{lorenzo.isella@isi.it}

\author[jrc,newcastle]{Yannis Drossinos\corref{cor}}
\ead{ioannis.drossinos@jrc.ec.europa.eu}

\cortext[cor]{Corresponding author.}
\fntext[present]{Present address: ISI Foundation, Turin 10133, Italy.}

\address[jrc]{European Commission, Joint Research Centre, I-21027
Ispra (VA), Italy}
\address[newcastle]{School of Mechanical \& Systems Engineering,
Newcastle University, Newcastle upon Tyne NE1 7RU, United Kingdom}


\begin{abstract}
A methodology to calculate the friction coefficient
of an aggregate in the continuum regime is proposed. The friction coefficient
and the monomer shielding factors, aggregate-average or individual, are related to the
molecule-aggregate collision rate that is obtained from the molecular diffusion equation
with an absorbing boundary condition on the aggregate surface.
Calculated friction coefficients of straight chains are in very good agreement
with previous results, suggesting that the friction coefficients
may be accurately calculated from the product of
the collision rate and an average momentum transfer, the
latter being independent of aggregate morphology.
Langevin-dynamics simulations show that the diffusive motion of straight-chain aggregates
may be described either by a monomer-dependent or an aggregate-average
random force, if the shielding factors are appropriately chosen.
\end{abstract}
%
%


\end{frontmatter}

\section{Introduction}
The hydrodynamic drag on a fractal aggregate suspended in
a viscous fluid determines many of its dynamical properties,
such as sedimentation, agglomeration, and mobility (diffusive, mechanical,
electrical)~\cite{friedlander-book}.
Fractal aggregates arise from the agglomeration of smaller,
primary spherules, hereafter called monomers, that do not coalesce, but rather retain
their identity in the resulting aggregate.
The calculation of the drag force can be a formidable task,
often requiring simplifying assumptions like spherical~\cite{castillo,vanni}
or ellipsoidal symmetry~\cite{shapiro}.

The friction coefficient of non-spherical aggregates in the continuum regime
(fluid mean free path much smaller than the monomer radius $R_1$)
has been calculated by treating the fractal aggregate as a porous
particle in a viscous creeping flow (Stokes flow)
of constant~\cite{shapiro}
or variable~\cite{castillo,vanni,veerapaneni} permeability.
Alternatively, Filippov~\cite{filippov} performed a multipole expansion of the
Stokes flow velocity with slip and non-slip boundary conditions on the
aggregate surface. Happel and Brenner~\cite{happel-brenner} calculated
the hydrodynamic forces on a collection of spheres by the method of reflections
obtaining a slowly converging series.
In these calculations the hydrodynamic force was obtained by integrating
the fluid stress tensor over the aggregate surface. For straight chains,
Dahneke~\cite{dahneke-friction}, based on an extrapolation of experimental data,
proposed an empirical correlation for the friction forces
parallel and perpendicular to the chain symmetry axis.

Herein, we propose a versatile, albeit approximate, procedure to determine the
friction coefficient of
an aggregate in creeping flows in the continuum regime.
We argue that the friction coefficient of a fractal-like
aggregate may be related to the ratio
of two molecular collision rates: the molecule-aggregate collision rate
and the molecule-monomer collision rate. These molecular collision rates are
calculated from the diffusion equation for the fluid density with appropriate
boundary conditions. Hence, we suggest that the friction coefficient of
fractal-like objects may be accurately (but approximately) calculated from
the solution of the Laplace equation (with appropriate boundary conditions)
for the fluid density, without requiring the 
solution of the Stokes equations for the fluid velocity.

The relationship between collision rates and friction coefficient allows us
to introduce the monomer shielding factor (either individual or aggregate-average)
that provides a measure of monomer shielding within an
aggregate. We apply the proposed methodology to calculate the friction
coefficients of straight-chain aggregates. The
calculated friction coefficients are compared to previous analytical results
and numerical results to justify the approximations made \textit{a posteriori},
and to validate the proposed methodology.
We remark, however, that the methodology is general enough to be applicable to
the calculation of friction coefficient of a general fractal-like aggregate.

Calculated shielding factors of individual monomers in straight chains composed
of $k=5,8$ monomers are used to obtain the chain diffusion coefficient
by Langevin-dynamics simulations. These coefficients are found to be in very good agreement
with the diffusion coefficients obtained from the ratio of molecular
collision rates.

Such Langevin simulations of aggregate formation and motion, which depend explicitly on the inter-monomer
interaction potential, offer a potentially useful tool to investigate the effect of inter-monomer
forces on agglomeration dynamics and the shape of the resulting aggregate. Previous
works have suggested, either experimentally or numerically, the importance of inter-particle
forces. For example, Chakrabarty el al.~\cite{LowFractalDimension} hypothesized
that electic forces were responsible for the unexpectedly low fractal 
dimension (between 1.2 and 1.5) of a minority of soot aggregates formed in a premixed ethene-oxygen flame.
di Stasio et al.~\cite{diStasio}, in their experimental and numerical studies
of the kinetics of agglomeration and growth of soot nanoparticles
in an ethylene-air diffusion flame, suggested that the number of collisions is
enhanced due to the attractive van der Waals forces. Kostoglou and co-workers~\cite{Kostoglou1,Kostoglou2}
studied theoretically the time-dependent evolution of aggregate morphology, as described by the aggregate
fractal dimension. Their analyses was based on a proposed constitutive law
for the agglomeration of two aggregates, a law that depends
on the inter-monomer (and hence inter-aggregate) interaction potential.
The Langevin simulations
described herein and in, e.g., Ref.~\cite{lorenzoLangevin}, could address these theoretical
and experimental observations by providing a technique to validate them.

\section{Molecule-aggregate collision rate and friction coefficient}
\label{sec:FrictionCoefficient}
\subsection{Methodology}
\label{sec:Methodology}

The motivation of our work stems from experimental measurements
of the mass transfer (attachment) coefficient of gas molecules to
nanoparticles of a variety of species, shapes, and sizes~\cite{siegmann,filippov-review-2}.
The rate of transfer of gas molecules of density $\rho$ to identical
aggregates composed of $k$ monomers ($k$-aggregates) of concentration $N_k$ is
\beq
\frac{d \rho}{dt} = - \tilde{K}_k N_k \rho \Comma
\label{eq:RateofTransfer}
\eeq
where $\tilde{K}_k$ is the total mass transfer coefficient of fluid
molecules per aggregate, namely $\tilde{K}_k = h_k S_k$ where $h_k$ is the
average heat transfer coefficient as usually defined in the
chemical engineering literature (per unit surface area) and $S_k$ the geometric surface
area of the aggregate. Equation~(\ref{eq:RateofTransfer}) is valid for
a molecule-aggregate sticking coefficient of unity.
Such mass transfer measurements may be performed by labelling atoms or molecules,
via, for example, radioactive labelling (epiphaniometer) or electrical charging (diffusion
charger), and subsequently detecting the atoms or molecules attached to the aggregate.
Siegmann and Siegmann~\cite{siegmann}, who measured total mass transfer coefficients
over a wide range of aggregate mobilities
via diffusive electrical charging, argued that the product of the
mass transfer coefficient times the aggregate (electrical) mobility $B_k$
is approximately constant
\beq
\tilde{K}_k \times B_k = \textrm{constant} \Comma
\label{eq:EmpiricalScalingLaw}
\eeq
where $B_k$, the aggregate mobility, is the ratio of aggregate terminal velocity to
the steady-state force causing it~\cite{hinds-book}.

Keller et al.~\cite{filippov-review-2} presented further experimental
measurements supporting the claim
that Eq.~(\ref{eq:EmpiricalScalingLaw}) hold (within experimental error) for
a variety of experimental conditions and aggregate shapes.
They suggested
that Eq.~(\ref{eq:EmpiricalScalingLaw}) constitutes an ``empirical scaling law''.
They justified it by arguing that both the mass transfer coefficient
and the aggregate mobility depend on the aggregate surface area exposed to the fluid.
They referred to the particle surface area upon which mass, momentum, and energy
transfer are dominant as their \textit{active} surface area.
Accordingly, the mass transfer coefficient is proportional
to the particle active surface area, whereas the mobility
is inversely proportional rendering their product a constant~\cite{siegmann, filippov-review-2}.
In particular, the total mass transfer coefficient $\tilde{K}_k$ to a sphere
in the continuum regime is proportional to its diameter $d_p$, whereas in the free
molecular regime it is proportional to its geometric surface area
($\sim d_p^2$).

If multiple scattering events are neglected (sticking coefficient of unity),
the attachment rate $\tilde{K}_k N_k$ is proportional to the molecule-aggregate
collision rate. For Stokes drag the mobility of a $k$-aggregate is
inversely proportional~\cite{hinds-book} to the aggregate friction
coefficient $f_k$. Equation~(\ref{eq:EmpiricalScalingLaw}), then, implies
that the collision rate  $K_k$ between a molecule and a $k$-aggregate is
proportional to the friction coefficient, $K_k = c f_k$. The proportionality
constant may be evaluated by considering, as experimental measurements
suggest, that Eq.~(\ref{eq:EmpiricalScalingLaw}) is approximately valid for $k=1$.
At steady state and for a spherical monomer, the integral molecule-monomer
collision rate $K_1$ is calculated from the
the gas diffusion equation with an absorbing
boundary condition on the monomer surface. Specifically,
\beq
K_{1} = \int_{S_1} \textbf{J}_1 \cdot d \textbf{S}
= \int_{S_1} \textbf{J}_1 \cdot \hat{\textbf{s}} \, d S
 = 4\, \pi \, D_{g} R_{1} \, \rho_{\infty} \Comma
\label{eq:MonomerCollisionRate}
\eeq
where the diffusive flux is $\textbf{J}_1 = -D_{g} \, {\nabla} \rho$
with $\rho(r) = \rho_{\infty} ( 1 - R_1/r)$,
$\rho_{\infty}$ is the gas-phase (fluid) density far away from the aggregate,
and $D_g$ the gas self-diffusion coefficient (see, for example, Ref.~\cite{friedlander-book}).
The monomer surface area is denoted by $S_1$, $\hat{\textbf{s}}$ is the unit vector orthogonal
to the monomer surface, and $d S$ the surface element ($dS = R_1^2 \sin \theta d \theta \, d \phi$).

The Stokes friction coefficient of an isolated monomer is
$f_1 = m_1 \beta_1 = 6 \pi \mu_g R_1$ with $\mu_g$ the fluid viscosity, leading to
the proportionality constant $c = 2 D_g \rho_{\infty}/(3 \mu_g)$ (for a similar calculation
of the particle active surface area in the continuum regime cf. Ref.~\cite{ActiveSurface}).
It is more convenient to eliminate the proportionality
constant by considering the drag force
$f_k$ relative to the total drag on $k$ isolated monomers $f_1$,
\beq
\frac{K_k}{k K_1} = \frac{f_k}{k f_1} \Teleia
\label{eq:RatioCollisionRates}
\eeq

Theoretical arguments provide partial support
of Eqs.~(\ref{eq:EmpiricalScalingLaw}), (\ref{eq:RatioCollisionRates}).
In kinetic theory the mean force exerted on a particle by fluid molecules is
the product of the molecule-particle collision rate times
the mean momentum transfer per collision~\cite{reif}. The main approximation of our work, and the approximation
implicit in Eqs.~(\ref{eq:EmpiricalScalingLaw}),  (\ref{eq:RatioCollisionRates}),
is that the dominant contribution to the friction
force arises from the collision rate. If the mean momentum transfer
is taken to be independent of aggregate morphology the ratio of the
friction coefficients  $f_k / (k f_1)$ becomes the appropriate
ratio of collision rates shown in Eq.~(\ref{eq:RatioCollisionRates}).
It is, thus, apparent that Eq.~(\ref{eq:RatioCollisionRates}) is approximate.
The  friction force arises from the momentum transfer
during molecule-particle collisions; even if the sticking probability is unity not all
collisions transfer the same momentum
since gas velocities are distributed according to a probability density function
that may, in principle, be calculated from the Boltzmann equation.

The Stokes friction coefficient of a $k$-aggregate is expressed
as $f_k = k m_1 \beta_k$ where $\beta_k$ is the average friction coefficient per unit monomer mass
and $k m_1$ the aggregate mass~\cite{lorenzoLangevin}. Equation~(\ref{eq:RatioCollisionRates}) then
leads to our main result that relates the friction coefficient of
a general-shaped aggregate to the ratio of two molecular collision rates
\beq
\frac{K_k}{k K_1} =
\frac{\beta_k}{\beta_1} \equiv \eta_k \Teleia
\label{eq:etak}
\eeq
The last equality defines the average monomer shielding factor $\eta_k$
as the ratio of the average friction coefficient of a monomer in an aggregate
to the friction coefficient of an isolated monomer. The shielding factor provides
a measure of the shielding of a monomer by other monomers in an aggregate; as such,
it depends on aggregate morphology.
It has been used to calculate corrections of the Stokes drag on a porous medium
consisting of identical non-interacting spheres~\cite{sonntag}, in modifications
of heat transfer to an aggregate due to monomer shielding~\cite{filippov-review},
and it has been related to the aggregate diffusion coefficient~\cite{lorenzoLangevin}.
The friction coefficient, calculated according to Eq.~(\ref{eq:etak}),
naturally leads to the Stokes-Einstein diffusion coefficient $D_k$ of a $k$-aggregate via
\beq
D_k = \frac{k_B T}{k m_1 \beta_k} = D_1 \, \frac{1}{k \eta_k} \Comma
\label{eq:DiffusionCoefficient}
\eeq
where the Stokes-Einstein monomer diffusion coefficient
is $D_1$[$=k_B T/(m_1 \beta_1$)]. Furthermore, the
aggregate mobility radius, defined as $D_k \equiv k_B T /(6 \pi \mu_g R_k)$, is
\beq
\frac{R_k}{R_1} = k \, \eta_k \Teleia
\label{eq:MobilityRadius}
\eeq
For the ideal aggregates defined in Ref.~\cite{lorenzoLangevin},
aggregates in the free-draining approximation whereby
the hydrodynamic forces on a monomer are independent
of its state of aggregation, $\eta_k = 1$. The shielding factor
also gives the ratio of a $k$-aggregate active surface area to the active
surface area of $k$ isolated monomers.

The dynamic shape factor of particles of arbitrary shape, a correction factor used to account for
the effect of shape on particle motion, is also related to the average
monomer shielding factor. The aggregate dynamic shape factor $\chi_s$ is defined by
\beq
\chi_k = \frac{f_k}{6 \pi \mu_g R_{eq}} \Comma
\eeq
where the $R_{eq}$ is the radius of the equivalent volume sphere,
$R_{eq} = k^{1/3} R_1$. Hence,
\beq
\chi_s = \eta_k \, k^{2/3} \Teleia
\label{eq:DynamicShapeFactor}
\eeq

As in the case of an isolated monomer, the integral collision rate
may be calculated from the fluid diffusion equation in steady state
[$\nabla^2 \rho(\textbf{r}) = 0$],
with an absorbing boundary condition on the aggregate surface
[$\rho (\textbf{r}_{\rm sur}) = 0$, neglect of multiple scattering events]
and constant density far away from the aggregate.
($\rho \to \rho_{\infty}$ for $|\textbf{r}| \to \infty$).
The diffusive flux $\textbf{J}_k = -D_{g} \, {\nabla} \rho$
yields the collision rate for a generic $k$-aggregate
\beq
K_{k} = \int_{S} \textbf{J}_k \cdot d \textbf{S}
= \int_{S} \textbf{J}_k \cdot \hat{\textbf{s}} \, d S \Comma
\label{eq:integration-for-collision-rate}
\eeq
where $S$ is the aggregate surface, $\hat{\textbf{s}}$ the unit vector orthogonal
to the aggregate surface, and $d S$ the surface element
[see, also, Eq.~(\ref{eq:MonomerCollisionRate})].
This calculation of the collision rate reflects the experimental
procedure to measure the attachment coefficient: the total mass transfer rate
is proportional to the integral collision rate.
Equation~(\ref{eq:integration-for-collision-rate})
highlights our approximations
in that the friction coefficient is approximately related to
a surface integral of the molecular diffusive flux instead of the stress tensor.
It is worthwhile noting that the linearized Bhatnagar-Gross-Krook equation (BGK),
a single relaxation-time approximation of the Boltzmann equation,
leads to a self-diffusion coefficient inversely proportional to the
BGK velocity-independent collision frequency~\cite{resibois}.

The proposed methodology suggests that the friction
coefficient of fractal-like aggregates may be accurately
calculated from the solution of the scalar Laplace
equation for the fluid density (with appropriate boundary
conditions) without solving the Stokes equations.
Once the fluid density
has been determined the collision rates may be easily calculated during 
post-processing via Eq.~(\ref{eq:integration-for-collision-rate}).
On the other hand, the Stokes equations (linear equations as
the Laplace equation) must be solved for the three components of the
fluid velocity to calculate the drag force. This replacement, in addition
to its theoretical implications,
has considerable computational advantages that depend primarily
on the symmetry of the aggregate and the dimensionality
of the system. Specifically, if axisymmetric aggregates are considered,
as in this work, the Stokes equations must be solved for the two components
of the drag force.
The Laplace equation, however, must be solved only once for the fluid density
in two dimensions.

\subsection{Friction coefficient of straight-chain aggregates}

Equations~(\ref{eq:etak}) and (\ref{eq:integration-for-collision-rate})
will be used to calculate average and individual monomer shielding factors
and the friction coefficient of straight monomer chains via 
the steady-state collision rate $K_{k}$ to a single aggregate.
We stress that Eq.~(\ref{eq:etak}) is general enough,
and easy to implement numerically, to be applicable to a generic fractal-like aggregate.
A straight-chain aggregate is cylindrically symmetric,
suggesting that a friction coefficient along the
axis orthogonal to the axis of symmetry $\beta^{\perp}_{k}$
and parallel to it $\beta_{k}^{\parallel}$ may be defined as
\beq
\f{K_{k}^{\perp}}{k K_{1}^{\perp}} =
\f{\beta_{k}^{\perp}}{\beta_{1}} \equiv \eta_k^{\perp} \Comma \quad
\f{K_{k}^{\parallel}}{k K_{1}^{\parallel}} =
\f{\beta_{k}^{\parallel}}{\beta_{1}} \equiv \eta_k^{\parallel} \Teleia
\label{eq:beta-orthogonal-and-perpendicular}
\eeq
The anisotropic collision rates $K_{k}^{\perp (\parallel)}$
are obtained by projecting the diffusive flux ${\bf J}_k$ parallel
and perpendicular to the symmetry axis, and integrating its absolute value
over the aggregate surface, cf. Eq.~(\ref{eq:integration-for-collision-rate}),
\begin{subequations}
\beq
K_k^{\parallel} \equiv \int_{S} \left | J_k^{\parallel} \right | \, d S
= \int_{S} \left | \textbf{J}_k \cdot \hat{\textbf{s}}_{\parallel} \right | d S
= \int_{S} \left | D_g \, \frac{\partial \rho}{\partial z} \right | d S
\Comma
\label{eq:AnisotropicFluxesParallel}
\eeq
\begin{eqnarray}
& K_k^{\perp} &  \equiv \int_{S} \left | J_k^{\perp} \right | \, d S
= \int_{S} \Big [ \left | \textbf{J}_k \cdot \hat{\textbf{s}}_{\perp, 1} \right |^2 +
\left | \textbf{J}_k \cdot \hat{\textbf{s}}_{\perp, 2} \right |^2 \Big ]^{1/2} d S \nonumber \\
& = & \int_{S} \left [ \left | \textbf{J}_k^x \right |^2 +
\left | \textbf{J}_k^y \right |^2 \right ]^{1/2} d S =
\int_{S} dS \left [ \left ( D_g \, \frac{\partial \rho}{\partial x} \right )^2 +
\left ( D_g \, \frac{\partial \rho}{\partial y} \right )^2 \right ]^{1/2},
\label{eq:AnisotropicFluxesPerpendicular}
\end{eqnarray}
\label{eq:AnisotropicFluxes}
\end{subequations}
where (in three dimensions and in a Cartesian co-ordinate system)
the unit vectors are $\hat{\textbf{s}}_{\parallel} = \hat{\textbf{s}}_z$,
$\hat{\textbf{s}}_{\perp,1} = \hat{\textbf{s}}_x$, and
$\hat{\textbf{s}}_{\perp,1} = \hat{\textbf{s}}_y$
with the $z$ coordinate along the symmetry axis and $x,y$ the co-ordinates
perpendicular to it. The absolute value
is necessary in Eq.~(\ref{eq:AnisotropicFluxes}) to ensure that the
collision rate is non-zero. As expected, Eq.~(\ref{eq:beta-orthogonal-and-perpendicular}) shows
that an isolated spherical monomer has only one isotropic friction
coefficient, $\beta_k^{\perp} = \beta_k^{\parallel} = \beta_1$.

According to Eq.~(\ref{eq:beta-orthogonal-and-perpendicular}) the parallel
and perpendicular fluxes to a spherical monomer should be calculated. Even
though a sphere is isotropic once a symmetry axis is randomly chosen the
two fluxes differ, $K_{1}^{\perp} \neq K_{1}^{\parallel}$. Let the symmetry axis be 
the $z$-axis. Since its choice is arbitrary (due to spherical symmetry)
\beq
\int_S \left | J_1^x \right | d S = \int_S \left | J_1^y \right | d S
= \int_S \left | J_1^z \right | d S = 2 \pi D_g R_1 \rho_{\infty} \Teleia
\label{eq:SphericalSymmetry}
\eeq
However, the perpendicular flux depends on the two fluxes in the
two perpendicular directions. Specifically,
the (local) molecular fluxes at the monomer surface are
$J_1^{\parallel} = J_1^z = J_1 \cos \theta $ and
$J_1^{\perp} =  ( | \textbf{J}_1^x |^2 + | \textbf{J}_1^y |^2 )^{1/2} = J_1 | \sin \theta| $
with $J_1 |_{r=R_1} = D_g \rho_{\infty} / R_1$, cf. Eq.~(\ref{eq:MonomerCollisionRate}).
Then, the analytical evaluation of the diffusional fluxes on
the surface of the monomer gives
\begin{subequations}
\beq
K_{1}^{\parallel} =  \int_{S} \left | J_1^{\parallel} \right | \, d S =
D_g \rho_{\infty} R_1 \int_0^{2 \pi} d \phi \int_0^{\pi} d \theta \left | \cos \theta \right | \sin \theta
=  2 \pi D_g R_1 \rho_{\infty} ,
\label{eq:k-par-andk-perp}
\eeq
\beq
K_{1}^{\perp} = \int_{S} \left | J_1^{\perp} \right | \, d S  =
D_g \rho_{\infty} R_1 \int_0^{2 \pi} d \phi \int_0^{\pi} d \theta \sin^2 \theta
=  \pi^2 D_g R_1 \rho_{\infty} \Teleia
\eeq
\end{subequations}
Note that, as expected from Eq.~(\ref{eq:AnisotropicFluxesPerpendicular}), $K_{1}^{\perp}\neq K_{1}^{\parallel}$
(in fact, $K_{1}^{\perp} > K_{1}^{\parallel}$), and that $K_1^{\parallel}$ is independent
of the choice of the symmetry axis, as shown in Eq.~(\ref{eq:SphericalSymmetry}).
We stress that even though the parallel and perpendicular diffusional fluxes as defined differ
for a spherical monomer, its diffusion coefficient is unique and it does not depend on
the choice of the symmetry axis.

We calculated the friction coefficients (average, and for motion parallel
and perpendicular to the symmetry axis) of straight chains consisting of up
to $k=64$ monomers. The finite-element software Comsol Multiphysics~\cite{comsol}
was used to solve the diffusion equation in cylindrical coordinates ($r,z$)
with an absorbing boundary condition on the chain surface. In this
co-ordinate system $J_k^{\parallel} = J_k^z$ and $J_k^{\perp} = J_k^r$.
We tested the mesh-independence of the solutions.
The size of the cylindrical computational
domain was at least two orders of magnitude
larger than the corresponding dimension of the chain
to ensure that the condition $\rho_{\infty}={\rm const}$
hold at the computational-domain boundaries.
The calculated diffusive flux to a single $k$-chain, and its 
components parallel and perpendicular to the
symmetry axis, was numerically integrated over the
aggregate surface to determine the molecule-aggregate collision rate.

Table~\ref{table-friction-coefficients} compares the friction coefficients
obtained via the ratio of collision rates to previous analytical and
numerical results.
The agreement is very good, justifying \textit{a posteriori} our main approximations.
For example, explicit calculation of the difference between our results and those
reported by Dahneke~\cite{dahneke-friction}
for chains composed of $k=2,3,4,5,8$ monomers shows that the maximum difference
is  $0.88$\%, $3.40$\%, and  $1.38$\% in $\eta_k$, $\eta_k^{\parallel}$, and
$\eta_k^{\perp}$, respectively.

\begin{table}[htb]
\begin{center}
\begin{tabular}{ccccc} \hline \hline
Friction & Filippov & Happel Brenner & Dahneke & Collision\\
coefficient & Ref.~\cite{filippov} & Ref.~\cite{happel-brenner} &  Ref.~\cite{dahneke-friction} & rate \\ \hline
$\eta_2 = \beta_{2}/\beta_{1}$ & & & $0.692$ & $0.694$ \\
$\eta_3 = \beta_{3}/\beta_{1}$ & & & $0.569$ & $0.574$ \\
$\eta_4 = \beta_{4}/\beta_{1}$ & & & $0.507$ & $0.507$ \\
$\eta_5 = \beta_{5}/\beta_{1}$ & & & $0.461$ & $0.463$ \\
$\eta_8 = \beta_{8}/\beta_{1}$ & & & $0.390$ & $0.389$ \\
& & & & \\
$\eta_2^{\parallel} = \beta^{\parallel}_{2}/\beta_{1}$ &  & $0.645 $ & $0.639$ & $0.633$ \\
$\eta_3^{\parallel} = \beta^{\parallel}_{3}/\beta_{1}$ & & &$0.511$ & $0.500$ \\
$\eta_4^{\parallel} = \beta^{\parallel}_{4}/\beta_{1}$ & & &$0.442$ & $0.430$ \\
$\eta_5^{\parallel} = \beta^{\parallel}_{5}/\beta_{1}$ & & &$0.397$ & $0.385$ \\
$\eta_8^{\parallel} = \beta^{\parallel}_{8}/\beta_{1}$ & & &  $0.324$ &$0.313$ \\
& & & & \\
$\eta_2^{\perp} = \beta^{\perp}_{2}/\beta_{1}$ & $0.726$ & $0.716$  & $0.719$ & $0.725$ \\
$\eta_3^{\perp} = \beta^{\perp}_{3}/\beta_{1}$ & $0.613$ & & $0.608$ & $0.612$ \\
$\eta_4^{\perp} = \beta^{\perp}_{4}/\beta_{1}$ & $0.550$ & & $0.545$ & $0.547$ \\
$\eta_5^{\perp} = \beta^{\perp}_{5}/\beta_{1}$ & $0.508$ & &$0.504$  & $0.503$  \\
$\eta_8^{\perp} = \beta^{\perp}_{8}/\beta_{1}$ & $0.435$ &   & $0.434$ & $0.428$ \\ \hline \hline
\end{tabular}
\caption{Comparison of calculated friction coefficients 
of straight chains of $k=2-8$ monomers (average, and for motion parallel and
perpendicular to the chain symmetry axis) with previous analytical and numerical results.
The values for $\eta_2^{\perp} - \eta_5^{\perp}$ in the second column where obtained from
reported values of the dynamic shape factor via Eq.~(\ref{eq:DynamicShapeFactor}).}
\label{table-friction-coefficients}
\end{center}
\end{table}

Filippov~\cite{filippov} calculated the friction
coefficient of non-overlapping spheres via a multipole expansion
of the Stokes flow velocity in a series of spherical
harmonics. The friction coefficient of a straight chain consisting of
$k=8$ monomers was reported to be $\beta_8^{\perp} / \beta_1 = 0.435$ (in our notation),
in good agreement with the calculated $0.428$, Table~\ref{table-friction-coefficients}.
The other values in the second column of Table~\ref{table-friction-coefficients} were obtained
from reported dynamic shape factors by inverting Eq.~(\ref{eq:DynamicShapeFactor}).

\begin{figure}[h]
\includegraphics[width=0.48\columnwidth]{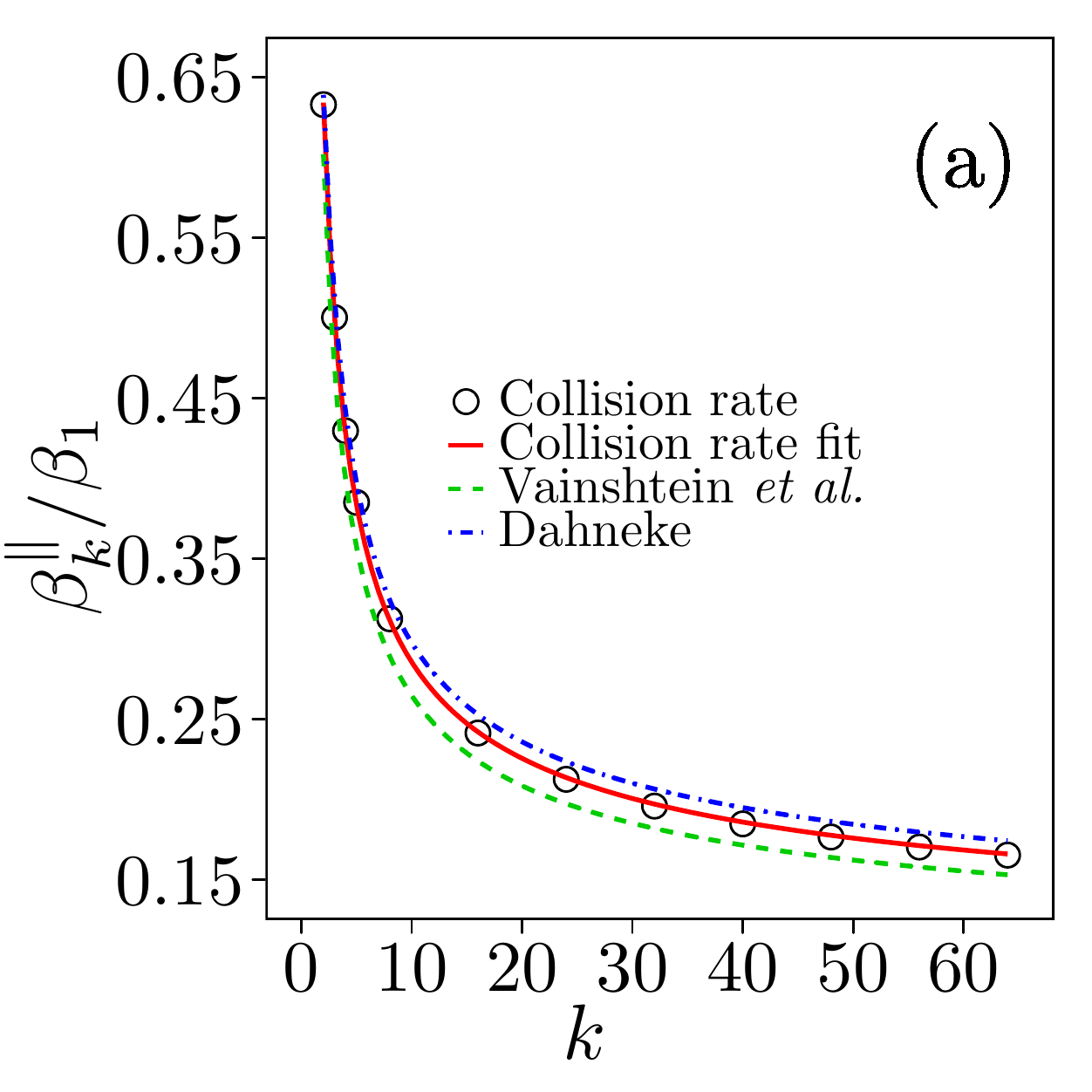}
\includegraphics[width=0.48\columnwidth]{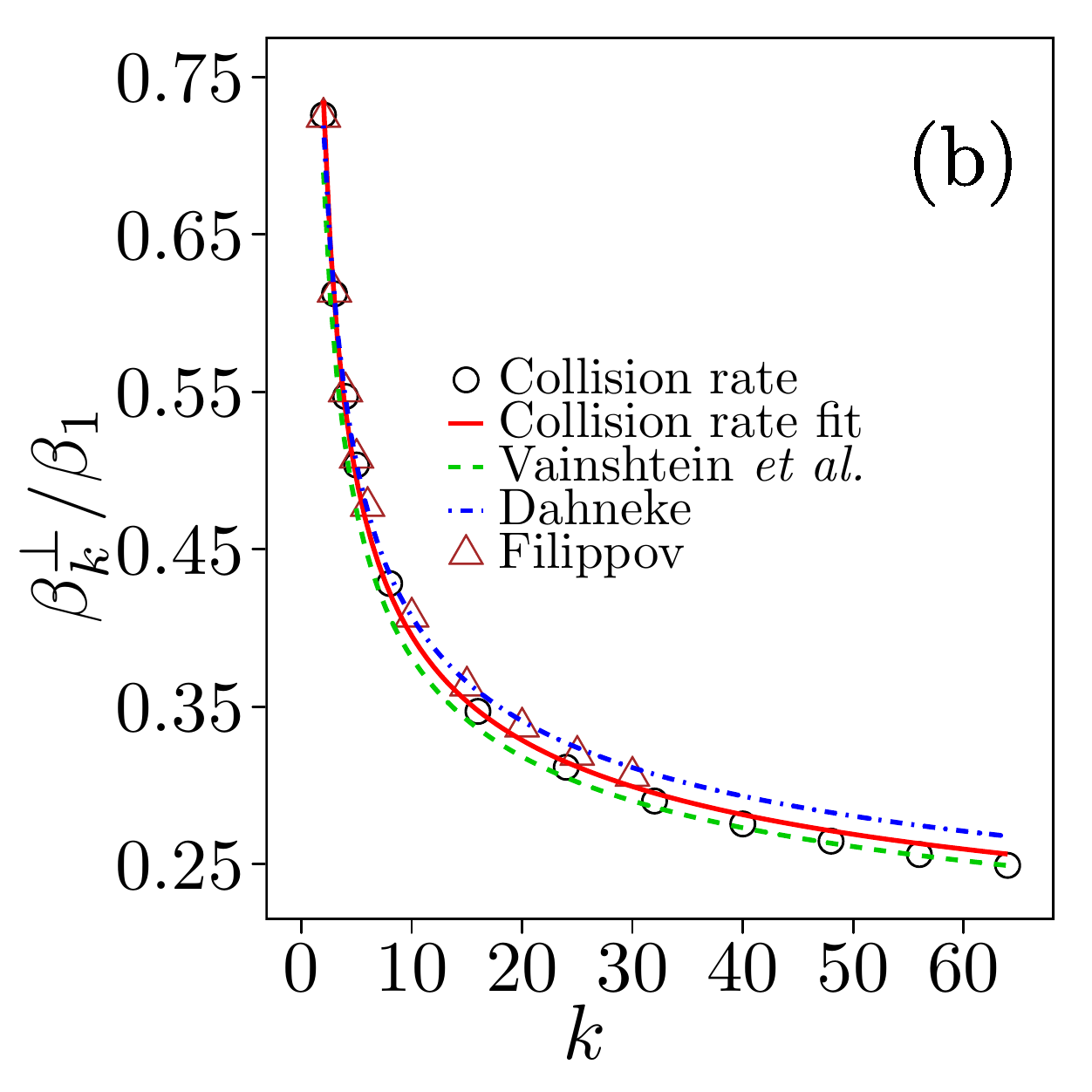}
\caption{Calculated friction coefficients of straight chains ($k = 2-64$)
for motion parallel (a) and perpendicular (b) to the chain symmetry axis
plotted against the number of monomers. Comparison with previous results.}
\label{fig:fit-friction}
\end{figure}

Dahneke~\cite{dahneke-friction}, based on an interpolation of experimental
measurements on chains with $k \leq 5$, provides extrapolation formulae for the
dimensionless drag forces
felt by a straight chain for motion parallel and perpendicular to the
symmetry axis. A straight chain composed of $k$ monomers was approximated as an
ellipsoid of aspect ratio $k$. In our notation the interpolation formulae read
\begin{eqnarray}
\eta_k^{{\parallel (\perp)}} & = & \f{\beta_k^{\parallel (\perp)}}{\beta_1} \nonumber \\
& = & \f{A_{\parallel (\perp) }(k^2-1)}{6 \pi k} \cdot
 \lsq \f{2(k^2 -1) \pm 1}{\sqrt{k^2-1}} \ln \lro k+ \sqrt{k^{2}-1} \rro
+ B_{\parallel (\perp)} k \rsq ^{-1} \Comma
\label{eq:fit-beta}
\end{eqnarray}
where the positive (negative) sign is chosen for the parallel (perpendicular)
coefficient. The formulae presented in Ref.~\cite{dahneke-friction}
map to ours via $\eta_k^{{\parallel \perp}} = c_{\parallel, \perp} (0) /(6 \pi k)$.
We fitted our calculated anisotropic friction coefficients to Eqs.~(\ref{eq:fit-beta}).
The fits were performed with the
Levenberg-Marquardt algorithm as implemented in the statistical language
R Minpack library~\cite{minpack}. We found $A_{\perp}=51.07$, $B_{\perp}=0.8637$,
$A_{\parallel}=27.45$, and $B_{\parallel}=-0.9386$ in
excellent agreement with the fit reported by Dahneke~\cite{dahneke-friction} (see, also, Fig.~\ref{fig:fit-friction}).

Vainshtein et al.~\cite{shapiro} performed an extensive analytical
study of the equivalent mobility radii of oblate and prolate spheroids.
Their results for non-porous impermeable prolate spheroids with aspect
ratio $k$ expressed in our notation are
\begin{subequations}
\beq
\eta_k^{\perp} = \f{\beta_{k}^{\perp}}{\beta_{1}} = \f{8}{3 k} \,
\lsq \f{k}{k^2 -1} + \f{2k^2 - 3}{(k^{2}-1)^{3/2}} \ln \lro k+ \sqrt{k^2-1} \rro \rsq^{-1} \Comma
\label{eq:beta-perp-shapiro}
\eeq
\beq
\eta_k^{\parallel} = \f{\beta_{k}^{\parallel}}{\beta_{1}}=\f{8}{3 k}
\lsq - \f{2k}{k^{2}-1} + \f{2k^{2}-1}{(k^{2}-1)^{3/2}} \ln \lro \f{k + \sqrt{k^{2}-1}}{k-\sqrt{k^{2}-1}}
\rro  \rsq ^{-1} \Teleia
\label{eq:beta-par-shapiro}
\eeq
\label{eq:Shapiro}
\end{subequations}
Equations~(\ref{eq:Shapiro}), expressed in terms of the mobility radius Eq.~(\ref{eq:MobilityRadius}),
are identical to the equations for the radius of the ``equivalent radius'' derived in
Happel \& Brenner~\cite{happel-brenner}.

In the limit of large straight $k$-chains Eqs.~(\ref{eq:fit-beta},~\ref{eq:Shapiro}) show that
their per-unit-mass friction coefficient tends to zero as $\beta_k \rightarrow 1/\ln(2k)$,
whereas the chain friction coefficient becomes infinitely large as $k \beta_k \rightarrow  k/\ln(2k)$.
The proportionality constant depends on the equations considered and on the
choice of the constants in Eq.~(\ref{eq:fit-beta}).
Moreover, Eqs.~(\ref{eq:Shapiro}) show that the parallel and perpendicular friction coefficients
tend to  $\beta_k^{\perp} \rightarrow 2 \beta_k^{\parallel}$ as $k \rightarrow \infty$,
whereas Eqs.~(\ref{eq:fit-beta}) predict 
$\beta_k^{\perp} \rightarrow \beta_k^{\parallel} \, A_{\perp} / A_{\parallel} $ with $ A_{\perp} / A_{\parallel} = 1.86$
[see, also, Eq.~(\ref{eq:RmOverRg})].

In Fig.~\ref{fig:fit-friction} we plot our numerical results (referred to as ``Collision
rate'', circles), the numerical fit based on Eqs.~(\ref{eq:fit-beta}) (``Collision rate fit'',
solid line), the anisotropic friction coefficients for an impermeable prolate spheroid
according to Vainstein et al.~\cite{shapiro}, Eqs.~(\ref{eq:Shapiro}), (``Vainstein \textit{ et al.}'', dashed line),
and the extrapolated results of Dahneke~\cite{dahneke-friction} (``Dahneke'', dot-dashed line). Friction
coefficients for motion parallel (left subfigure) and perpendicular to the chain symmetry axis
are presented.  In the right subfigure, squares denote results reported by Filippov~\cite{filippov} ($k=8$)
or derived from reported dynamic shape factors.
As previously remarked
the agreement of our calculated friction coefficients to those predicted
by Eqs.~(\ref{eq:fit-beta}) and (\ref{eq:Shapiro}) is very good.

\begin{figure}
\includegraphics[width=0.48\columnwidth]{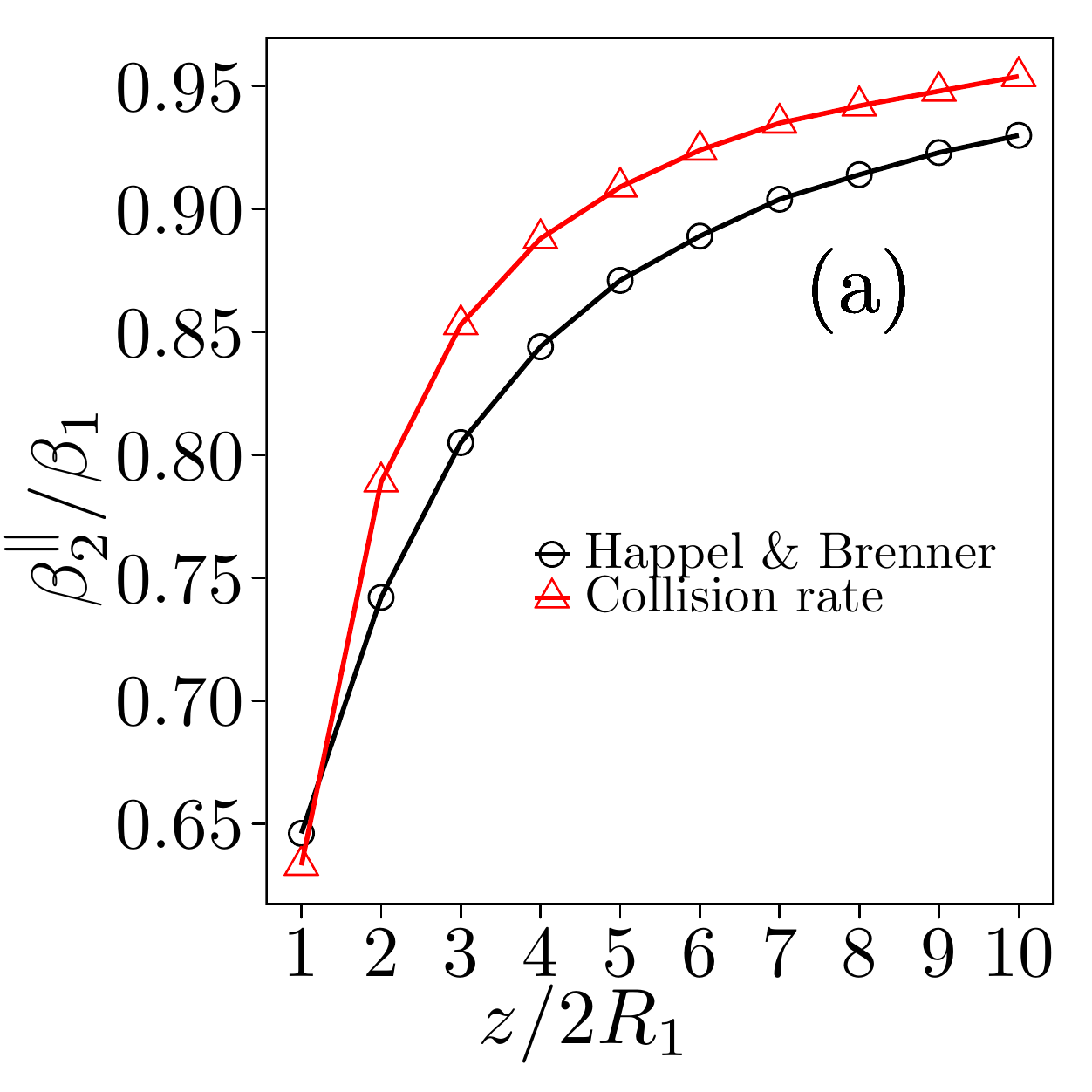}
\includegraphics[width=0.48\columnwidth]{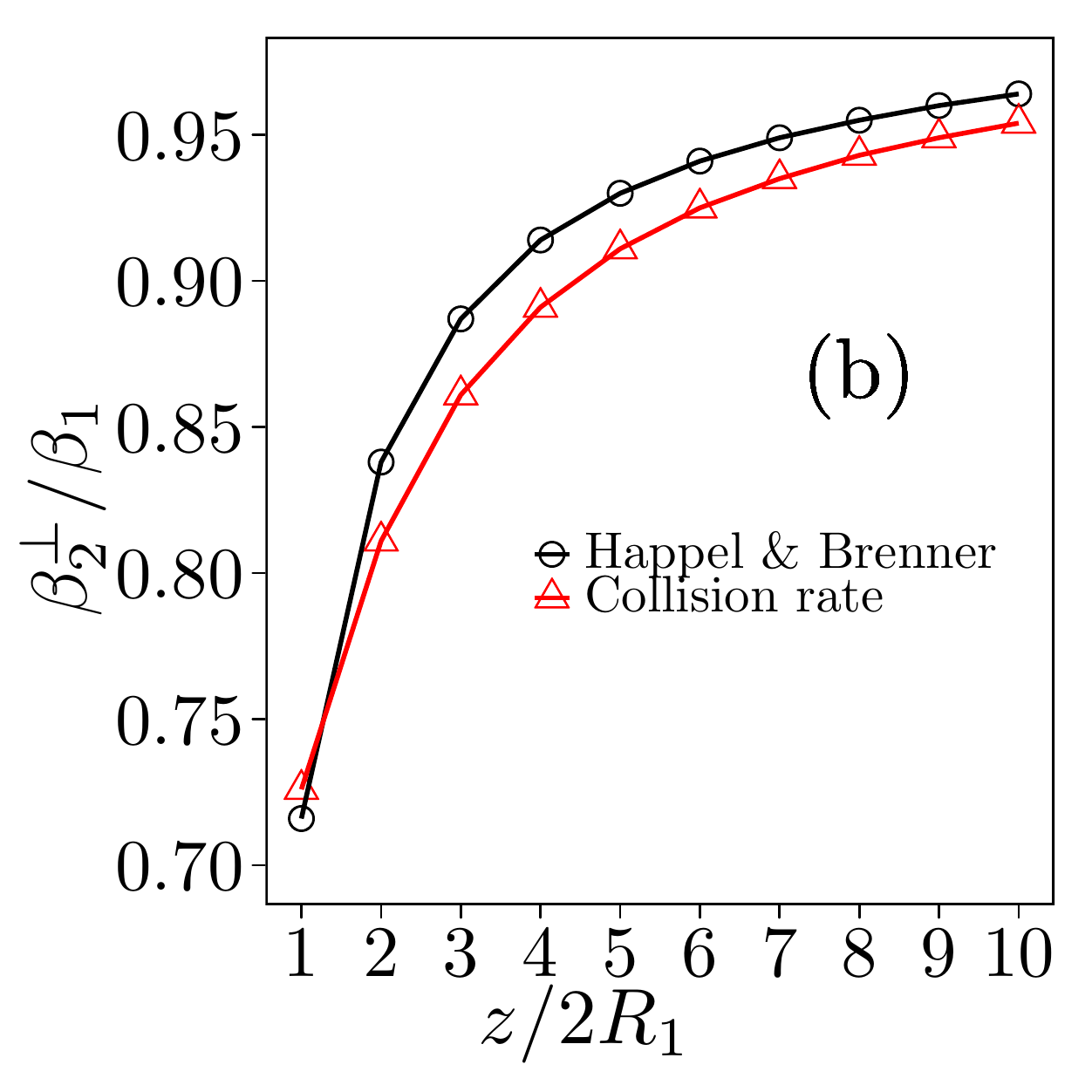}
\caption{Calculated parallel (a) and perpendicular (b) friction coefficients of
two equal-sized, non-rotating spheres as a function of their
center-of-mass distance $z$. Comparison with the results of Happer \& Brenner~\cite{happel-brenner}.}
\label{fig:Two-spheres-distance}
\end{figure}

Happel \& Brenner~\cite{happel-brenner} presented a comprehensive
analysis of the resistance coefficients of two equal-size spheres
moving along the line of their centers or perpendicular to it.
They solved the Stokes equations via the method of reflection.
Their results for the friction coefficients of
two equal-size non-rotating spheres as a function
of their separation are compared to ours in Fig.~\ref{fig:Two-spheres-distance}.
In particular, for two touching spheres their results $\beta_2^{\parallel} / \beta_1 = 0.645$
and $\beta_2^{\perp} / \beta_1 = 0.716$ compare favourably to ours,
$0.633$ and $0.725$, respectively.
We remark that the comparison is best for touching spheres.

Table~\ref{table:DynamicShapeFactors} presents calculated dynamic shape factors for
the aggregates considered in our simulations. The shape factors where calculated via
Eq.~(\ref{eq:DynamicShapeFactor}) by substituting the appropriate shielding factor
($\eta_k, \eta_k^{\parallel}, \eta_k^{\perp}$). The last column presents the
results of Ref.~\cite{filippov} determined from a multipole expansion of the Stokes
velocity field.

\begin{table}[htb]
\begin{center}
\begin{tabular}{ccccc} \hline \hline
Number of & Average dynamic & Parallel dynamic & \multicolumn{2}{c}{Perpendicular dynamic} \\
monomers & shape factor ($\chi_k$) & shape factor ($\chi_k^{\parallel}$) & \multicolumn{2}{c}{shape factor ($\chi_k^{\perp}$)} \\
& Collision & Collision & Collision & Filippov \\
& rate & rate & rate & Ref.~\cite{filippov} \\ \hline
$2$ & $1.102$ & $1.005$ & $1.152$ & $1.15$ \\
$3$ & $1.194$ & $1.041$ & $1.274$ & $1.276$ \\
$4$ & $1.278$ & $1.083$ & $1.379$ & $1.386$ \\
$5$ & $1.354$ & $1.126$ & $1.472$ & $1.485$ \\
$8$ & $1.556$ & $1.250$ & $1.713$ & \\
$16$ & $1.975$ & $1.533$ & $2.203$ & \\
$24$ & $2.312$ & $1.769$ & $2.592$ & \\
$32$ & $2.601$ & $1.975$ & $2.924$ & \\
$40$ & $2.862$ & $2.162$ & $3.223$ & \\
$48$ & $3.100$ & $2.334$ & $3.495$ & \\
$56$ & $3.322$ & $2.495$ & $3.748$ & \\
$64$ & $3.531$ & $2.647$ & $3.987$ & \\ \hline \hline
\end{tabular}
\caption{Average dynamic shape factor ($\chi_k$) of straight chains,
and their dynamic shape factor for motion parallel ($\chi_k^{\parallel}$) and
perpendicular ($\chi_k^{\perp}$) to their symmetry axis. The last column
reports results of Ref.~\cite{filippov}.}
\label{table:DynamicShapeFactors}
\end{center}
\end{table}

\begin{figure}[hbt]
\begin{center}
\includegraphics[width=0.50\columnwidth]{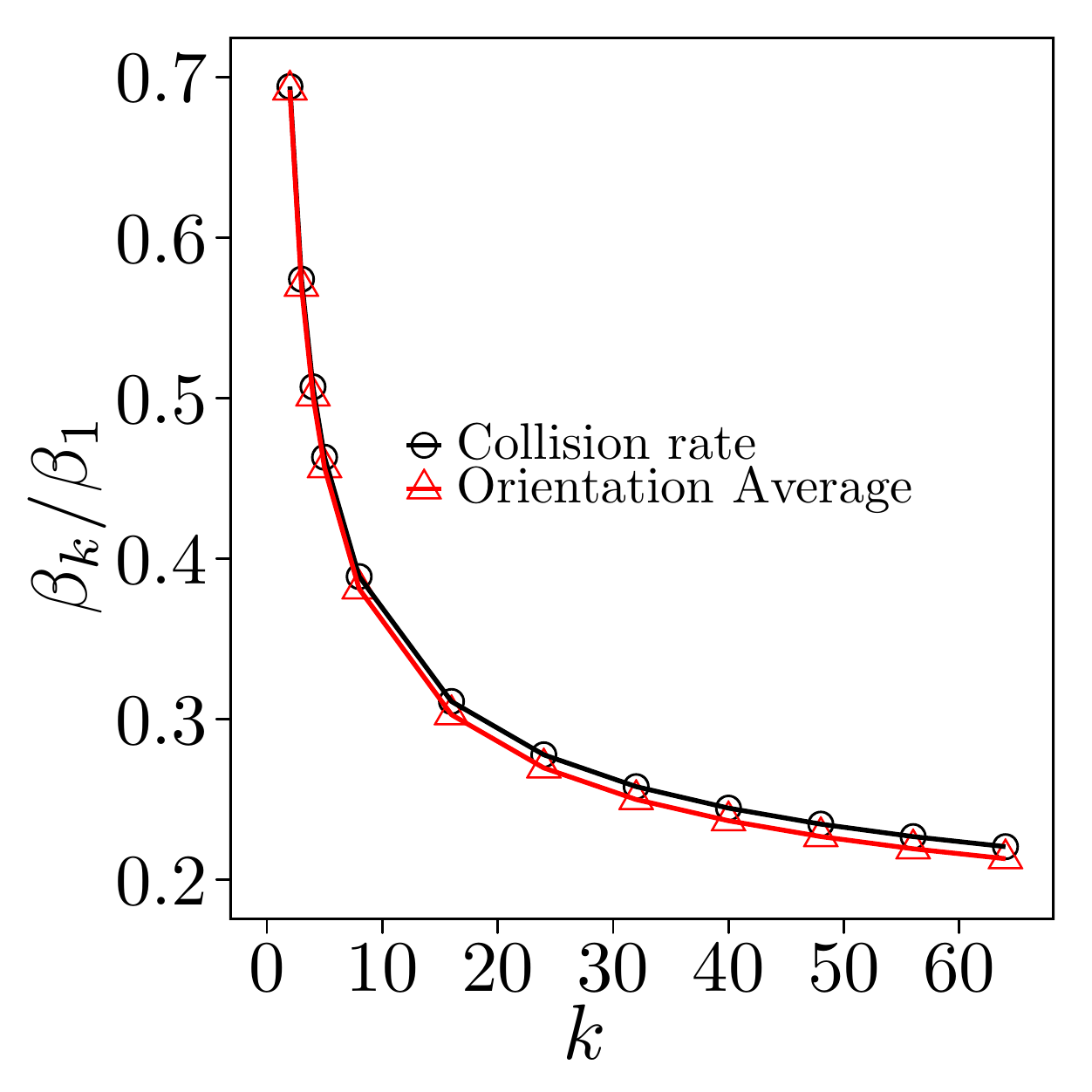}
\caption{Isotropic straight-chain shielding factors
$\eta_k = \beta_{k}/\beta_1$ calculated via the total collision rates from the diffusion simulations (diamonds)
and from anisotropic collision rates and Eq.~(\ref{eq:RandomOrientation}) (triangles) ($k= 2-64$).}
\label{fig:fit-friction-isotropic}
\end{center}
\end{figure}

Dahneke~\cite{dahneke-friction} also argued that for a large ensemble
of identical straight chains undergoing random Brownian rotations
the orientation-averaged friction coefficient would be
\beq
\beta_k = \frac{3 \beta_{k}^{\parallel} \beta_{k}^{\perp}}{\beta_{k}^{\perp} + 2 \beta_{k}^{\parallel}} \Teleia
\label{eq:RandomOrientation}
\eeq
Figure~\ref{fig:fit-friction-isotropic} compares $\beta_{k}/\beta_1$ evaluated
using $\bpar /\beta_1$ and $\bperp / \beta_1$ from the ratio of the 
appropriate collision rates Eq.~(\ref{eq:AnisotropicFluxes}) and Eq.~(\ref{eq:RandomOrientation}) (crosses) to the
direct calculation of $\beta_{k}/ \beta_1$ (diamonds) via Eq.~(\ref{eq:etak}) and (\ref{eq:integration-for-collision-rate}).
The agreement is very satisfactory.

\begin{figure}[hbt]
\begin{center}
\includegraphics[width=0.48\columnwidth]{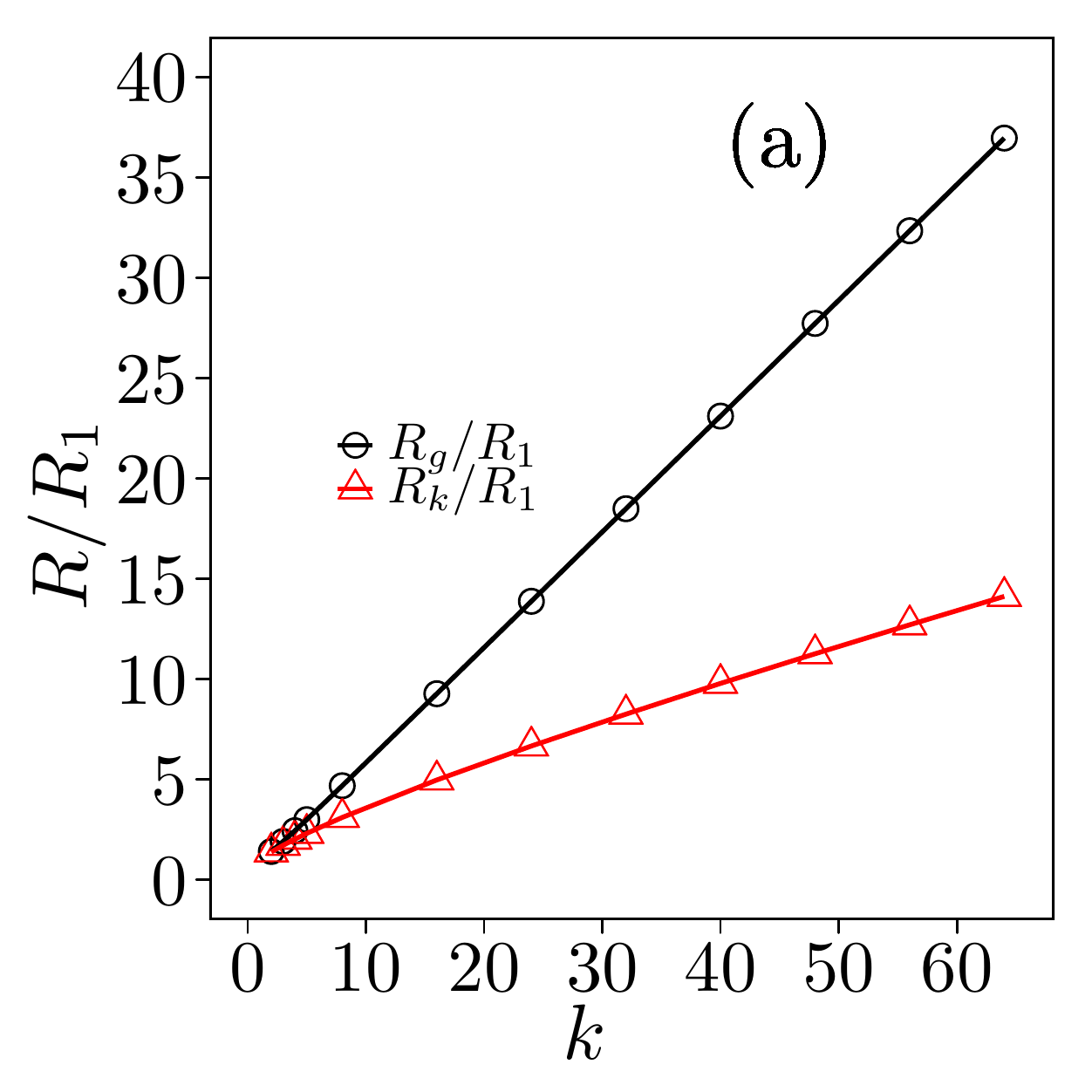}
\includegraphics[width=0.48\columnwidth]{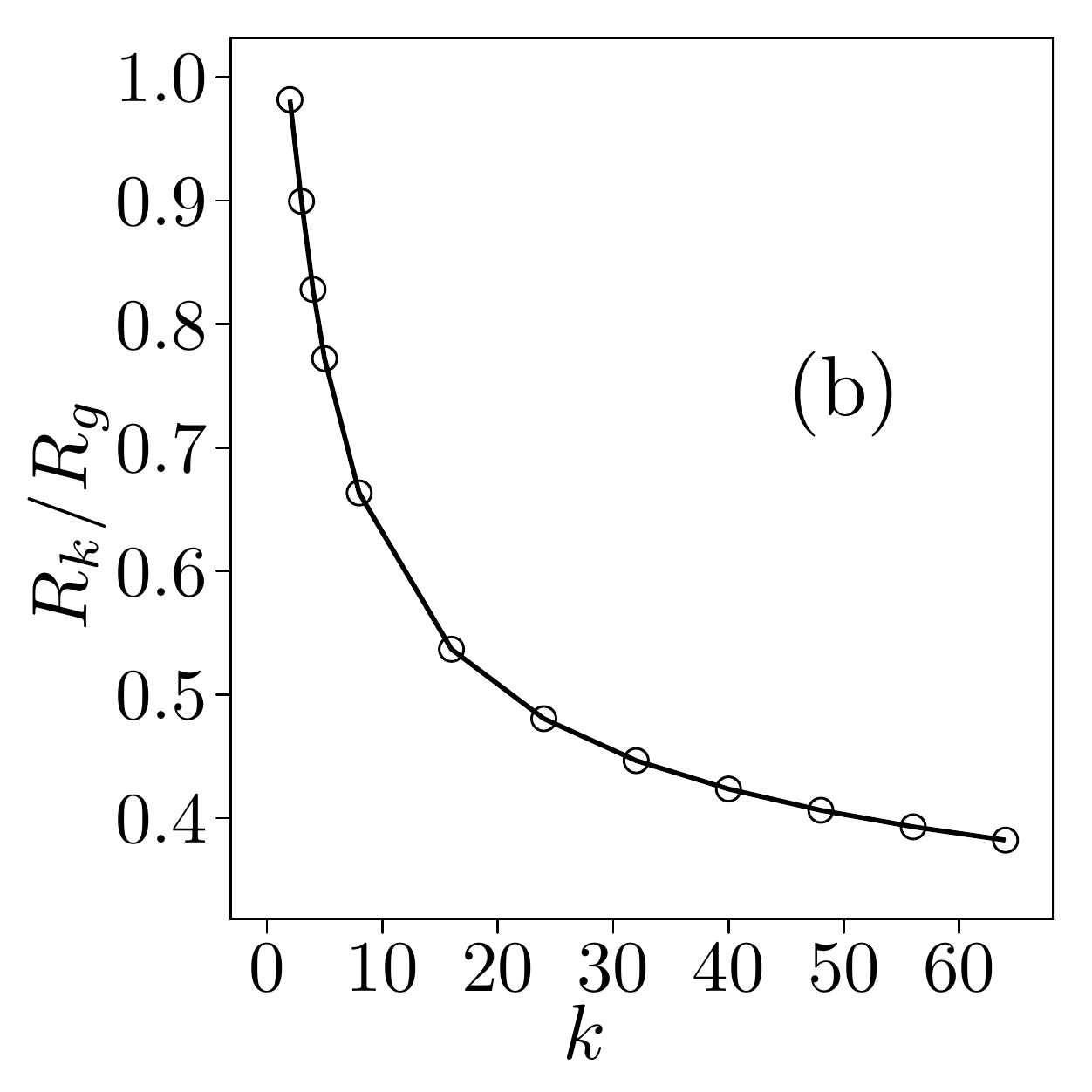}
\caption{Left: Dimensionless straight-chain mobility diameter $R_k/R_1$ (triangles)
and radius of gyration $R_g/R_1$ (circles) plotted against number of monomers ($k= 2-64$).
Right: Ratio of the mobility radius to the radius of gyration of straight chains as a function
monomer number ($k= 2-64$).}
\label{fig:MobilityRadius}
\end{center}
\end{figure}

The mobility radius $R_k$ of straight chains calculated by Eq.~(\ref{eq:MobilityRadius})
is shown in Fig.~\ref{fig:MobilityRadius}(a). As expected, the mobility radii of
chains composed of shielded monomers are considerably smaller than the mobility radii of ideal chains
[for which $R_k \sim k$, Eq.~(\ref{eq:MobilityRadius})]. In addition,
their radius of gyration $R_g$ is plotted. It is easy to show that the radius of gyration of
straight $k$-chains is
\beq
R_g^2 = \frac{1}{3} \, R_1^2 \left ( k^2 - 1 \right ) + R_1^2 \Teleia
\label{eq:RadiusOfGyration}
\eeq
The last term $R_1^2$ is an additional term to the usual definition of the
radius of gyration to ensure that the radius of gyration of a monomer is non-zero.
The choice shown in Eq.~(\ref{eq:RadiusOfGyration}) corresponds to
the geometric radius of a sphere~\cite{filippov}; alternatively, the radius of gyration of a
sphere ($R_1 \sqrt{3/5}$) has been used~\cite{lorenzoLangevin}.

The ratio of the mobility diameter to the radius of gyration is shown in
Fig.~\ref{fig:MobilityRadius}(b). Note that for straight chains the two diameters
differ significantly, as also noted in Refs.~\cite{filippov,filippov-review}.
They argued that the usual assumption $R_k \sim R_g$ may be justified for
dense fractal-like aggregates with fractal dimension greater than $2$, but the
approximate equality fails for smaller dimensions, in agreement with our results
shown in Fig~\ref{fig:MobilityRadius}(b) for straight chains.
For large straight chains in the limit $k \rightarrow \infty$ Eqs.~(\ref{eq:fit-beta}) (with the
constants determined from the fit of the collision-rate friction coefficients) and
Eqs.~(\ref{eq:RandomOrientation},~\ref{eq:RadiusOfGyration}) show that the
ratio of the mobility radius to the radius of gyration tends to
\beq
\frac{R_k}{R_g} = 1.05 \, \frac{\sqrt{3}}{\ln(2k)} \quad \textrm{as} \quad k \rightarrow \infty \Comma
\label{eq:RmOverRg}
\eeq
in good agreement with the slender-body theory prediction~\cite{filippov,Batchelor70}
\beq
\frac{R_k}{R_g} = \frac{\sqrt{3}}{\ln(2k)} + O \left ( \frac{1}{\ln^2(2k)} \right ) \Teleia
\label{eq:Batchelor}
\eeq
Equations~(\ref{eq:Shapiro}) reproduce the slender-body theory result.

The connection between collision rates and the average monomer shielding factor
may be refined to obtain the shielding factor of a monomer within
an aggregate. The intra-chain (isotropic, per unit monomer mass) friction
coefficient $\beta_{k}^{(i)}$ of monomer $i$ in a $k$-aggregate is defined by
\beq
\f{K_k^{(i)}}{K_1} = \f{\beta_k^{(i)}}{\beta_1} \equiv \eta_{k}^{(i)} \Comma
\label{eq:friction-monomer}
\eeq
where $K_{k}^{(i)}$ is the steady-state molecular collision rate of the $i$th
monomer in the $k$-chain. It is easy to show that
\beq
\eta_k = \frac{1}{k} \sum_{i=1}^k \eta_k^{(i)} \quad \Longrightarrow \quad
f_k = \sum_{i=1}^k f_k^{(i)} = \sum_{i=1}^k \eta_k^{(i)} f_1 \Comma
\label{eq:IndividualShieldingfactors}
\eeq
where $f_k$ is the total aggregate friction coefficient, $f_k^{(i)}$
the total friction coefficient of the $i$th monomer, and $f_1=m_1 \beta_1$
the monomer friction coefficient.
Equation~(\ref{eq:friction-monomer}), as Eq.~(\ref{eq:etak}), can be applied to aggregates of arbitrary shape.
Table~\ref{table:etak} presents position-dependent monomer shielding factors in $k=5,8$ straight chains.

\begin{table}[hbt]
\begin{center}
\begin{tabular}{ccc|cccc} \hline \hline
$\eta_5^{(1)}$ & $\eta_5^{(2)}$ & $\eta_5^{(3)}$ &
$\eta_8^{(1)}$ & $\eta_8^{(2)}$ & $\eta_8^{(3)}$ & $\eta_8^{(4)}$ \\ \hline
$ 0.597$ & $ 0.379$ & $ 0.364$ &
$ 0.565$ & $ 0.350$ & $ 0.325$ & $ 0.317$ \\ \hline \hline
\end{tabular}
\caption{Monomer shielding factors in two straight chains, $k=5,8$.
The superscript denotes the relative position of the monomer in the chain,
the subscript the total number of monomers.}
\label{table:etak}
\end{center}
\end{table}

\section{Langevin dynamics of straight-chain aggregates}

Langevin simulations have been used extensively to investigate
aggregate collisional dynamics, see, for example, Refs.~\cite{lorenzoLangevin,mountain},
and in particular the diffusive motion of aggregates. The Brownian
motion of an aggregate may be described by modelling the Brownian motion of a
set of interacting monomers held together by strong monomer-monomer interaction forces.
A coupled set of Langevin equations, each one for a monomer within
the aggregate, is solved to determine the diffusive motion of the whole aggregate.
The stochastic properties of random force, which describes the effect of molecular
collisions with the aggregate, are usually determined by assuming that
the Fluctuation Dissipation Theorem (FDT) hold for each monomer.
Accordingly, the random force acting on each monomer is determined from its friction coefficient.
The monomer friction coefficient is usually assumed to be independent of aggregate
morphology and equal to the average monomer friction coefficient $\beta_k$.
In particular, in the free draining approximation (i.e., for ideal clusters) the monomer friction
coefficient is taken to be the friction coefficient of an isolated monomer~\cite{lorenzoLangevin}.
In the continuum regime, these assumptions yield the aggregate diffusion
coefficient $D_k$ as given in Eq.~(\ref{eq:DiffusionCoefficient}).

The previously described determination of the intra-chain shielding factor allows
the introduction of a monomer-dependent random force. In the following we use
these intra-chain shielding factors to calculate the diffusion coefficient of straight chains
from the Langevin dynamics of each monomer in the chain.
%
%
%
The Langevin equations of motion of the $i$th monomer in a $k$-chain is
\beq
m_1 \ddot{\bf r}_i = {\bf F}_i-\beta_k^{(i)} m_1 \dot{\bf r}_i+{\bf W}_i(t) \Comma
\label{eq:Langevin}
\eeq
where ${\bf r}_i$ is the monomer position, ${\bf F}_i$ the external force acting on it, and
${\bf W}_i$ the random force that models the effect of molecule-monomer collisions.
The stochastic properties of the noise are usually taken to be delta-correlated in time
and space ($j,j'=x,y,z$; monomers are identified by $i,i'$)
\beq A_{\perp} / A_{\parallel}
\langle  W^{j}_i(t) \rangle = 0 \ ,
\langle W^{j}_i(t) W_{i'}^{j'}(t') \rangle  = \Gamma_{i} \delta_{ii'} \delta_{jj'} \delta(t-t') \Teleia
\label{eq:gaussian_noise}
\eeq
The monomer-dependent noise strength $\Gamma_{i}$ is determined from the FDT to be
\beq
\Gamma_{i} = 2 \beta_k^{(i)} m_1 k_B T \Comma
\label{eq:FDTmonomer}
\eeq
explicitly depending on the monomer shielding factor. The corresponding noise strength
in Langevin simulations where the average monomer friction coefficient $\beta_k$ is used becomes
$\Gamma = 2 \beta_k m_1 k_B T$, i.e., it is independent of the monomer position in the chain.

\begin{figure}[htp]
\begin{center}
\includegraphics[width=0.75\columnwidth]{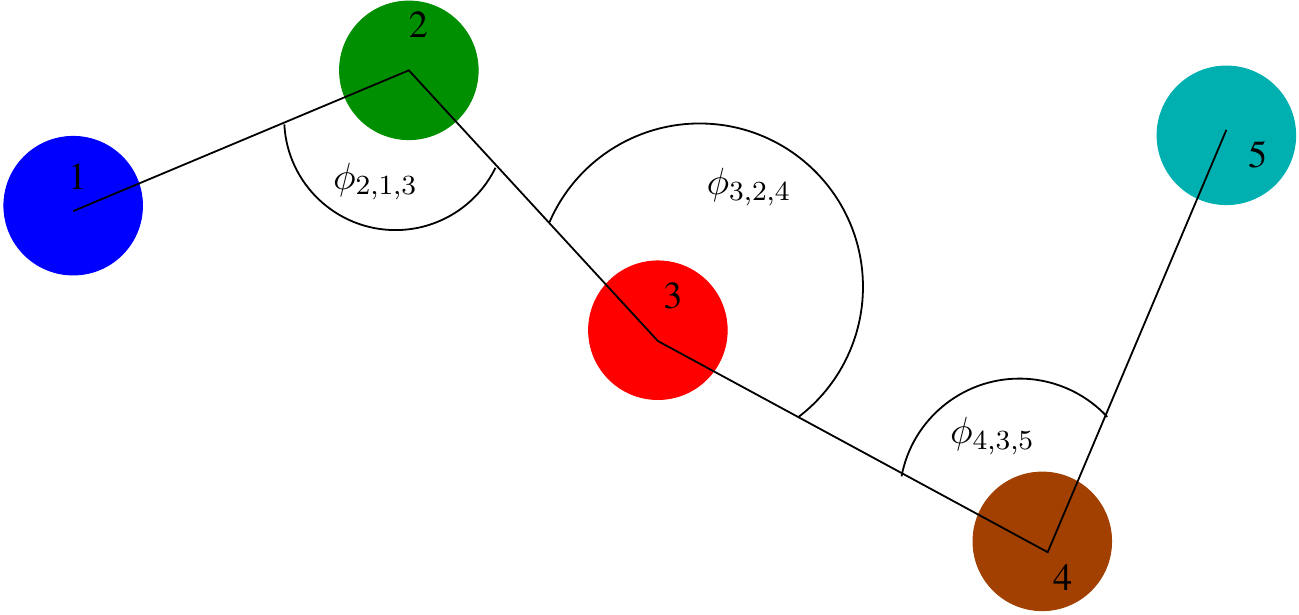}
\caption{Definition of the angles used in the angular monomer-monomer
interaction potential $U_i^{\textrm{ang}}$ for a linear chain of $k=5$ monomers.}
\label{espresso-angle-application-chain}
\end{center}
\end{figure}

The force in Eq.~(\ref{eq:Langevin}) models monomer-monomer interactions.
It will be taken to be conservative
\beq
{\bf F}_i=- \mbox{\boldmath$\nabla$}_{{\bf r}_i} U_i \Comma
\label{eq:potential_pairwise}
\eeq
where $U_i$ is the total intermonomer potential the $i$th monomer feels.
The monomer potential will be chosen to ensure the stiffness of the straight chain.
We split potential into a radial part $U_i^{\textrm{rad}}$ and
an angular part $U_i^{\textrm{ang}}$. The radial potential fixes the distance between any
two first-neighbor monomers along the chain, and the angle-dependent potential $U_{i}^{\textrm{ang}}$
prevents the chain from bending. The radial part is taken to be pairwise additive
\beq
U_i^{\textrm{rad}} = \s_{j\neq i}^k u_{ij}(r_{ij}) \Comma
\label{eq:pairwise_pot_explicit}
\eeq
where $r_{ij}=|{\bf r}_i-{\bf r}_j|$. In our simulations the radial, two-body potential
$u_{ij}(r_{ij})$ was chosen to
be the pair-wise additive model potential described in Ref.~\cite{lorenzoLangevin}.
The angular potential $U_i^{\textrm{ang}}$ depends on the angle $\phi_{i,i-1,i+1}$
formed by the three consecutive monomers ($i-1$, $i$, $i+1$) as follows
\beq
U_i^{\textrm{ang}}=\Omega\lsq 1- \cos(\phi_{i,i-1,i+1} -\pi) \rsq, \  i = 2, \ldots , k-1 \Comma
\label{eq:espresso-bond-angle}
\eeq
where $\Omega$ is a constant that determines the bending stiffness of the chain.
Figure~\ref{espresso-angle-application-chain} illustrates
the definition of $\phi_{i,i-1,i+1}$ within a $k=5$ chain.
The monomer-monomer potential described by Eqs.~(\ref{eq:pairwise_pot_explicit}) and
(\ref{eq:espresso-bond-angle}) introduces a penalty function for any
radial or angular deformation of the linear chain to ensure that the chain remain
straight and rigid during its Brownian motion.

We performed the numerical simulations of the mobility of straight chains
with the molecular dynamics package \esp~\cite{espresso}, modified to
allow for a monomer-dependent random force.
The depth of the radial inter-monomer potential was chosen deep enough
to ensure that monomers remain bound. Similarly, the strength of
the angular potential $\Omega$ was chosen to ensure that the linear chain did not bend during
the simulation ($\Omega = 500 k_B T$).

The chain diffusion coefficient $D_k^{LD}$ was estimated from the time dependence of the chain
centre-of-mass mean-square displacement
\beq
\lim_{t\to \infty} \langle \delta r^{2}_{CM}(t) \rangle =  6 D_k^{LD} t \Comma
\label{eq:DiffusionCoefficientLangevin}
\eeq
where brackets denote an ensemble average.
We followed the motion of the aggregate centre-of-mass along $3200$
trajectories up to $\beta_1 t=100$ for chains of $k=5,8$.
The chain was placed in the middle of a box of size $L$ with
$L/(2 R_1) = 10000$.
The box size was large enough so that no overall displacement of the aggregate
larger than $L/2$ was observed.
Details of similar simulations are described in Ref.~\cite{lorenzoLangevin}.

\begin{figure}[htb]
\centering
\includegraphics[width=0.50\columnwidth]{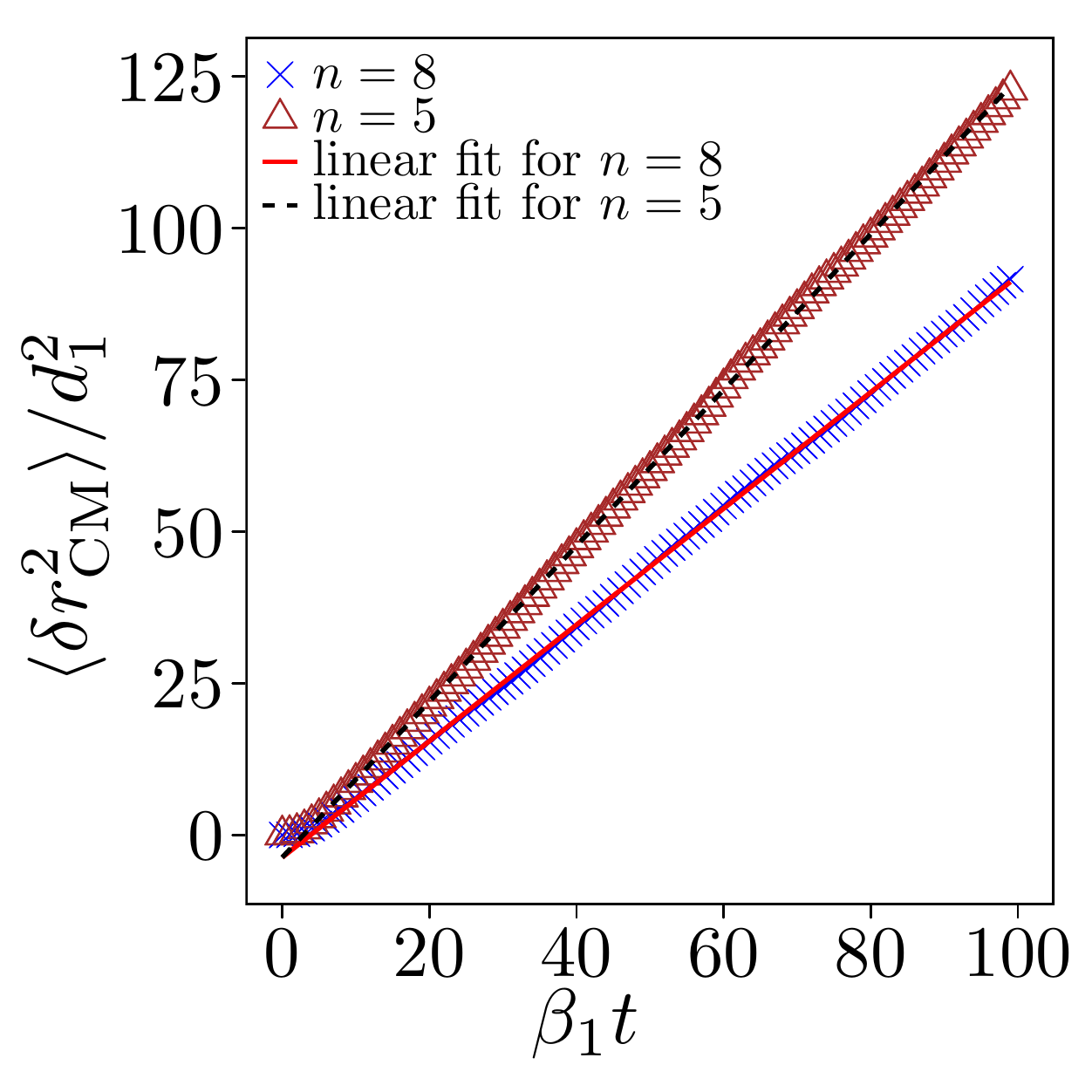}
\caption{Normalized mean-square displacements of the center-of-mass of
straight chains consisting of $k=5,8$ monomers ($d_1$ is the monomer diameter).
The linear fits were performed for $\beta_{1}t \ge 3$.}
\label{fig:fit-diffusion-langevin}
\end{figure}

The dimensionless chain center-of-mass displacements (scaled by the monomer diameter $d_1$) are shown in
Fig.~\ref{fig:fit-diffusion-langevin} as a function of dimensionless time
for the two straight chains.
The numerical fit of the linear dependence of
$\langle\delta  r^{2}_{\rm CM}\rangle$ on time was performed for
$\beta_{1}t\ge 3$. We find $D_{5}^{LD}/D_{1}=0.428$ and $D_{8}^{LD}/D_{1}=0.319$,
were $D_1$ is the monomer diffusion coefficient.
These values are within $1\%$ of the diffusion coefficient
$D_{5}/D_{1}=0.432$ and $D_{8}/D_{1}=0.321$ obtained from
Eq.~(\ref{eq:DiffusionCoefficient}) and the shielding factors reported
in Table~\ref{table-friction-coefficients}, i.e., via the average
monomer shielding factor calculated from the total collision rates.
Since the diffusion coefficient is inversely proportional to
the friction coefficient these results confirm numerically
that the aggregate friction coefficient is the sum of the
fiction coefficient of the shielded monomers, Eq.~(\ref{eq:IndividualShieldingfactors}).

Therefore, the Langevin simulations provide additional support to
the proposed methodology whereby the diffusion coefficient of straight chains
may be accurately calculated
via the ratio of molecular collision rates. Furthermore,
they suggest that calculations of $k$-aggregate diffusion
may be performed by two different, but equivalent, approaches.
An average monomer friction coefficient $\beta_k = \eta_k \beta_1$ may be used on each monomer
in the aggregate, as in usual Langevin simulations via, e.g., \esp.
The FDT is assumed to hold for an average monomer and the noise strength is the same for all monomers
$\Gamma = 2 \beta_k m_1 k_BT$. Alternatively, individual monomer friction coefficients $\beta_k^{(i)}$
($i=1, \ldots k$) are introduced, and the complete set of monomer-dependent
friction coefficients $\{ \beta_k^{(i)} \}$ (or, equivalently, $\{ \eta_k^{(i)} \}$)
has to be specified. The FDT is, again, assumed to hold
for each monomer, but the noise strength becomes monomer-dependent
$\Gamma_{i} = 2 \beta_k^{(i)} m_1 k_BT$, Eq.~(\ref{eq:Langevin}). Both
descriptions of aggregate Brownian motion are consistent
because they yield the same diffusion coefficient $D_k$.

\section{Conclusions}
We developed a methodology to calculate approximately, albeit accurately, the friction and
diffusion coefficients of generic fractal-like aggregates via the ratio
of molecule-aggregate collision rates. The collision rates
are obtained from numerical simulations of the molecular diffusion equation
with an absorbing boundary condition on the aggregate surface.
The methodology was validated for straight chains via comparison of friction
coefficients, isotropic and anisotropic, with analytical calculations and extrapolations
of experimental measurements.
Since the numerical solution of the diffusion equation (a scalar Laplace equation)
is often more feasible
than sophisticated treatments of the Stokes equations
relatively accurate insights on aggregate mobility and monomer shielding factors may
be gained from collision-rate simulations.

In addition to the average shielding factor of a monomer in an aggregate,
the collision rate on each monomer was used to define a
monomer-dependent intra-chain friction coefficient.
Langevin-dynamics simulations of the diffusive motion of
straight chains were performed to argue that within our approximation
the Fluctuation Dissipation Theorem could be equally well applied on the whole aggregate by
considering an average monomer or on each monomer,
as long as the appropriate shielding factors, and consequently
the friction coefficients, are used. The diffusion coefficients
of straight chains composed of $k=5,8$ monomers calculated by Langevin
simulations were found to be in good agreement with the diffusion coefficients calculated via
the ratio of the molecular collision rates.
The Langevin simulations
depended explicitly on the inter-monomer interaction potential, thereby providing
a useful technique to investigate the effect of inter-particle forces on
agglomeration dynamics and on the shape of the resulting aggregates.

\section*{Acknowledgements}

We thank Anastasios Melas for useful discussion and additional collision-rate simulations.


\begin{thebibliography}{99}

\bibitem{friedlander-book} S.K. Friedlander, Smoke, Dust and Haze,
Oxford University Press, New York, 2000.

\bibitem{castillo} P.L. Garcia-Ybarra, J.L. Castillo, D.E. Rosner,
J. Aerosol Sci. 37 (2006) 413.

\bibitem{vanni} M. Vanni, Chem. Eng. Sci. 55 (2000) 685.

\bibitem{shapiro} P. Vainshtein, M. Shapiro, C. Gutfinger, J. Aerosol Sci. 35
(2004) 383.


\bibitem{veerapaneni} S. Veerapaneni, M.R. Wiesner, J. Colloid Inter. Sci. 177 (1996) 45.

\bibitem{filippov} A.V. Filippov, J. Colloid Interface Sci. 229 (2000) 184.

\bibitem{happel-brenner} J. Happel, H. Brenner, Low Reynolds number hydrodynamics,
2nd edn., Kluwer Academic Press, Dordrecht, Holland, 1991.

\bibitem{dahneke-friction} B. Dahneke, Aerosol Sci. Technol. 1 (1982) 179.

\bibitem{LowFractalDimension} R.K. Chakrabarty, H. Moosm\"uller, W. Patrick Arnott, M.A. Garro, G. Tian, J.G. Slowik,
E.S. Cross, J.-H. Han, P. Davidovits, T.B. Onasch, D.R. Worsnop, Phys. Rev. Lett. 102 (2009) 235504; 104 (2010) 119602;
M. Sander, R.I.A. Patterson, A. Raj, M. Kraft, \textit{ibid.} 104 (2010) 119601;

\bibitem{diStasio} S. Di Stasio, A.G. Konstandopoulos, M. Kostoglou, J. Colloid Inter. Sci. 247 (2002) 33.

\bibitem{Kostoglou1} M. Kostoglou, A.G. Konstandopoulos, J. Aerosol Sci. 32 (2001) 1399.

\bibitem{Kostoglou2} M. Kostoglou, A.G. Konstandopoulos, S.K. Friedlander, J. Aerosol Sci. 37 (2006) 1102.

\bibitem{lorenzoLangevin} L. Isella, Y. Drossinos, Phys. Rev E 82 (2010) 011404.

\bibitem{siegmann} K. Siegmann, H.C. Siegmann, Society of Automotive Engineers
Technical Paper (2000) 2000-01-1995.

\bibitem{filippov-review-2} A. Keller, K. Siegmann, A.C. Siegmann, A. Filippov,
J. Vacuum Sci. Technol. A 19 (2001) 1.

\bibitem{hinds-book} W.C. Hinds, Aerosol Technology, 2nd edn., Wiley \& Sons, New York, 1999.

\bibitem{ActiveSurface} W.A. Heitbrink, D.E. Evans, B.K. Ku, A.D. Maynard, T.J. Slavin,
T.M. Peters, J. Occup. Environ. Hyg. 6 (2009) 19.

\bibitem{reif} F. Reif, Fundamentals of Statistical and Thermal Physics,
McGraw-Hill, New York, 1965.


\bibitem{sonntag} R.C. Sonntag, W.B. Russel, J. Colloid Inter. Sci. 115 (1987) 378.

\bibitem{filippov-review} A.V. Filippov, M. Zurita, D.E. Rosner, J. Colloid Interface
Sci. 229 (2000) 261.

\bibitem{resibois} P. R\'esibois, M. de Leener, Classical Kinetic Theory of Fluids,
John Wiley \& Sons, New York, 1977.

\bibitem{minpack} T.V. Elzhov and K.M. Mullen, \textit{Minpack.lm: R interface for least squares optimization library},
\url{http://cran.r-project.org/web/packages/minpack.lm/index.html} (2008).

\bibitem{comsol} Comsol Multiphysics, Chemical Engineering Module, version 3.5,
\url{http://www.comsol.com/products/chem/} (2008).



\bibitem{Batchelor70} G.K. Batchelor, J. Fluid Mechanics 44 (1970) 419.

\bibitem{mountain} R.D. Mountain, G.W. Mulholland, H. Baum, J.  Colloid Interface Sci. 114 (1986) 67.

\bibitem{espresso} H.J. Limbach, A. Arnold, B.A. Mann, C. Holm, Computer Physics
Communications 174 (2006) 704.

\end{thebibliography}
\end{document}